# Brain volume: An important determinant of functional outcome after acute ischemic stroke

Markus D. Schirmer, PhD[a,b,c,*]; Kathleen L. Donahue, BS[a]; Marco J. Nardin, BA[a]; Adrian V. Dalca, PhD[b,d]; Anne-Katrin Giese, MD[a]; Mark R. Etherton, MD, PhD[a]; Steven J. T. Mocking, MS[d]; Elissa C. McIntosh, MA[d]; John W. Cole, MD[e]; Lukas Holmegaard, MD, MSc[f]; Katarina Jood, MD, PhD[f]; Jordi Jimenez-Conde, MD, PhD[g]; Steven J. Kittner, MD, MPH[e]; Robin Lemmens, MD, PhD[h]; James F. Meschia, MD[i]; Jonathan Rosand, MD, MSc[a,d,j]; Jaume Roquer, MD, PhD[g]; Tatjana Rundek, MD, PhD[k]; Ralph L. Sacco MD, MS[k]; Reinhold Schmidt, MD[l]; Pankaj Sharma, MD, PhD[m]; Agnieszka Slowik, MD, PhD[n]; Tara M. Stanne, PhD[f]; Achala Vagal, MD, MS[o]; Johan Wasselius, MD, PhD[p]; Daniel Woo, MD, MS[q]; Stephen Bevan, PhD[r]; Laura Heitsch, MD[s]; Chia-Ling Phuah, MD, MMSc[t]; Daniel Strbian MD, PhD[u]; Turgut Tatlisumak, MD, PhD[v]; Christopher R. Levi, M.B., B.S[w]; John Attia, MD, PhD[x]; Patrick F. McArdle, PhD[y]; Bradford B. Worrall, MD, MSc[z]; Ona Wu, PhD[d]; Christina Jern, MD, PhD[f]; Arne Lindgren, MD, PhD[ab,ac]; Jane Maguire, PhD[ad]; Vincent Thijs, MD, PhD[ae]; Natalia S. Rost, MD, MPH[a]; *on behalf of the MRI-GENIE and GISCOME Investigators and the International Stroke Genetics Consortium*

[a]Stroke Division & Massachusetts General Hospital, J. Philip Kistler Stroke Research Center, Harvard Medical School, Boston, USA

[b]Computer Science and Artificial Intelligence Lab, Massachusetts Institute of Technology, Boston, USA

[c]Department of Population Health Sciences, German Centre for Neurodegenerative Diseases (DZNE), Germany

[d]Athinoula A. Martinos Center for Biomedical Imaging, Department of Radiology, Massachusetts General Hospital, Charlestown, MA, USA

[e]Department of Neurology, University of Maryland School of Medicine and Veterans Affairs Maryland Health Care System, Baltimore, MD, USA.

[f]Institute of Biomedicine, the Sahlgrenska Academy at University of Gothenburg, Gothenburg, Sweden.

[g]Department of Neurology, Neurovascular Research Group (NEUVAS), IMIM-Hospital del Mar (Institut Hospital del Mar d'Investigacions Mèdiques), Universitat Autonoma de Barcelona, Barcelona, Spain.

[h]KU Leuven - University of Leuven, Department of Neurosciences, Experimental Neurology and Leuven Research Institute for Neuroscience and Disease (LIND), Leuven, Belgium; VIB, Vesalius Research Center, Laboratory of Neurobiology, University Hospitals Leuven, Department of Neurology, Leuven, Belgium

[i]Department of Neurology, Mayo Clinic, Jacksonville, FL, USA.

[j]Center for Genomic Medicine, Massachusetts General Hospital, Boston, MA, USA.

[k]Department of Neurology and Evelyn F. McKnight Brain Institute, Miller School of Medicine, University of Miami, Miami, FL, USA.

[l]Department of Neurology, Clinical Division of Neurogeriatrics, Medical University Graz, Graz, Austria.

[m]Institute of Cardiovascular Research, Royal Holloway University of London (ICR2UL), Egham, UK St Peter's and Ashford Hospitals, UK.

[n]Department of Neurology, Jagiellonian University Medical College, Krakow, Poland.

[o]Department of Radiology, University of Cincinnati College of Medicine, Cincinnati, OH, USA.

[p]Department of Clinical Sciences Lund, Radiology, Lund University, Lund, Sweden; Department of Radiology, Neuroradiology, Skåne University Hospital, Malmö, Sweden.

[q]Department of Neurology and Rehabilitation Medicine, University of Cincinnati College of Medicine, Cincinnati, OH, USA.

[r]School of Life Science, University of Lincoln, Lincoln, UK.

[s]Division of Emergency Medicine, Washington University School of Medicine, St Louis, MO.

[t]Department of Neurology, Washington University School of Medicine & Barnes-Jewish Hospital, St Louis, MO.

[u]Division of Neurocritical Care & Emergency Neurology, Department of Neurology, Helsinki University Central Hospital, Helsinki, Finland.

[v]Department of Clinical Neuroscience, Institute of Neuroscience and Physiology, Sahlgrenska Academy at University of Gothenburg, Gothenburg, Sweden; Department of Neurology, Sahlgrenska University Hospital, Gothenburg, Sweden.

[w]School of Medicine and Public Health, University of Newcastle, Newcastle, New South Wales, Australia; Department of Neurology, John Hunter Hospital, Newcastle, NSW, Australia.

[x]Hunter Medical Research Institute, Newcastle, New South Wales, Australia; School of Medicine and Public Health, University of Newcastle, NSW, Australia.

[y]Division of Endocrinology, Diabetes and Nutrition, Department of Medicine, University of Maryland School of Medicine, Baltimore, MD, USA.

[z]Departments of Neurology and Public Health Sciences, University of Virginia, Charlottesville, VA, USA.

[ab]Department of Neurology and Rehabilitation Medicine, Skåne University Hospital, Lund, Sweden.

[ac]Department of Clinical Sciences Lund, Neurology, Lund University, Lund, Sweden

[ad]University of Technology Sydney, Sydney, Australia

[ae]Stroke Division, Florey Institute of Neuroscience and Mental Health, Heidelberg, Australia and Department of Neurology, Austin Health, Heidelberg, Australia.

* Corresponding author
Markus D. Schirmer
J. Philip Kistler Stroke Research Center,
Massachusetts General Hospital,
175 Cambridge Street, Suite 300,
Boston, MA 02114, USA
Tel: 617 643 7597
Fax: 617 643 3939
ORCID: 0000-0001-9561-0239

Email: mschirmer1@mgh.harvard.edu

**Acknowledgements:**
This project has received funding from the European Union's Horizon 2020 research and innovation programme under the Marie Sklodowska-Curie grant agreement No 753896. Major support for this study was provided by the MRI-GENIE study (NIH-NINDS R01NS086905, Rost PI), the Swedish Research Council (C.J.), the Swedish Heart and Lung Foundation (C.J.), and the Swedish State under the ALF agreement (C.J.).

**Disclosures**:

Dr. Schirmer is in part supported by Horizon 2020 grant agreement No 753896.

Ms. Donahue reports no disclosures.

Mr. Nardin reports no disclosures.

Dr. Dalca is supported by NIH 1R21AG050122.

Dr. Giese reports no disclosures.

Dr. Etherton is supported in part by the AHA Clinical Scientist Training Program (17CPOST33680102).

Mr. Mocking reports no disclosures.

Ms. McIntosh reports no disclosures.

Dr. Cole is partially supported by National Institutes of Health (NIH) grants R01-NS100178 and R01-NS105150, the US Department of Veterans Affairs, the American Heart Association (AHA) Cardiovascular Genome-Phenome Study (Grant-15GPSPG23770000), and the AHA-Bayer Discovery Grant (Grant-17IBDG33700328).

Dr. Holmegaard reports no disclosures.

Dr. Jood reports no disclosures.

Dr. Jimenez-Conde in part supported by Fondos de Investigación Sanitaria ISC III (PI10/02064), (PI15/00451); and Fondos FEDER/EDRF Spanish stroke research network INVICTUS+ (RD16/0019/0021).

Dr Kittner is partially supported by NIH grant, R01 NS086905-01.

Dr. Lemmens is a senior clinical investigator of FWO Flanders.

Dr. Meschia is co-Principal Investigator for the CREST-2 trial (U01 NS080168) and receives some support for the CREST-H study (R01 NS097876). He also serves as chair for the NeuroNEXT DSMB.

Dr. Rosand is supported by grants from the US National Institutes of Health and the OneMind Foundation and the Henry and Allison McCance Center for Brain Health.

Dr. Roquer is in part supported by Spain's Ministry of Health (Ministerio de Sanidad y Consumo, Instituto de Salud Carlos III FEDER, RD12/0042/0020).

Dr. Rundek is supported by the grants from the NINDS R37 NS 29993; R01NS040807; and NIH/NCATS Miami Clinical and Translational Science Institute U54TR002736/ KL2TR002737.

Dr. Sacco is supported by the grants from the NINDS R37 NS 29993; R01NS040807; and NIH/NCATS Miami Clinical and Translational Science Institute U54TR002736/ KL2TR002737.

Dr. Schmidt reports no disclosures.

Dr. Sharma reports no disclosures.

Dr. Slowik reports no disclosures.

Dr. Stanne reports no disclosures.

Dr. Vagal is supported by R01 NINDS NS100417, R01 NINDS NS30678, NIH/NINDS 1U01NS100699, research grant Cerovenus, GE Healthcare research grant.

Dr. Wasselius reports no disclosures.

Dr. Woo is in part supported by NIH funds.

Dr. Bevan reports no disclosures.

Dr. Heitsch reports no disclosures.

Dr. Phuah reports no disclosures.

Dr. Strbian received funding from the Helsinki University Central Hospital governmental subsidiary funds for clinical research.

Dr. Tatlisumak reports grants from Helsinki University Central Hospital, grants from University of Gothenburg, grants from Sahlgrenska University Hospital, grants from Sigrid Juselius Foundation, during the conduct of the study; grants and personal fees from Boehringer Ingelheim, personal fees from Bayer, grants from BrainsGate, grants from Bayer, grants from Pfizer, personal fees from Lumosa Pharm, grants from Portola Pharm, outside the submitted work; In addition, Dr. Tatlisumak has a patent use of a mast cell activation or degranulation blocking agent in the manufacture of a medicament for the treatment of a patient subjected to thrombolyses. Patent no: US8163734. Filed: February 13, 2004. Issued: April 24, 2012.

Mr. Levi reports no disclosures.

Dr. Attia reports no disclosures.

Dr. McArdle is in part supported by NIH NINDS (R01 NS100178, R01 NS105150).

Dr. Worrall is in part supported by U10 NS086513/U24 NS107222 and U01 NS069208. Dr. Worrall is the Deputy Editor of the journal Neurology.

Dr. Wu is in part supported by NIH (P50NS051343, R01NS082285, R01NS086905) and National Institute of Biomedical Imaging and Bioengineering (P41EB015896).

Dr. Jern reports no disclosures.

Dr. Lindgren is supported by Region Skåne, Lund University, the Swedish Heart and Lung Foundation, the Freemasons Lodge of Instruction Eos Lund, Skåne University Hospital, the Foundation of Färs&Frosta—one of Sparbanken Skåne's ownership Foundations, and the Swedish Stroke Association.

Dr. Maguire reports no disclosures.

Dr. Thijs has been a consultant for Bayer, Boehringer Ingelheim, BMS/Pfizer and Amgen; is in part supported by the Victorian Government and in particular the funding from the Operational Infrastructure Support Grant.

Dr. Rost is in part supported by NIH-NINDS (R01NS086905 and R01NS082285).

This work was presented as a poster at Brain & Brain PET 2019 in Yokohama, Japan (July 4-7, 2019).

# Abstract

**Objective**: To determine whether brain volume is associated with functional outcome after acute ischemic stroke (AIS).

**Methods**: We analyzed cross-sectional data of the multi-site, international hospital-based MRI-GENetics Interface Exploration (MRI-GENIE) study (July 1, 2014- March 16, 2019) with clinical brain magnetic resonance imaging (MRI) obtained on admission for index stroke and functional outcome assessment. Post-stroke outcome was determined using the modified Rankin Scale (mRS) score (0-6; 0: asymptomatic; 6 death) recorded between 60-190 days after stroke. Demographics and other clinical variables including acute stroke severity (measured as National Institutes of Health Stroke Scale score), vascular risk factors, and etiologic stroke subtypes (Causative Classification of Stroke) were recorded during index admission.

**Results**: Utilizing the data from 912 acute ischemic stroke (AIS) patients (65±15 years of age, 58% male, 57% history of smoking, and 65% hypertensive) in a generalized linear model, brain volume (per 155.1cm$^3$) was associated with age (β -0.3 (per 14.4 years)), male sex (β 1.0) and prior stroke (β -0.2). In the multivariable outcome model, brain volume was an independent predictor of mRS (β -0.233), with reduced odds of worse long-term functional outcomes (OR: 0.8, 95% CI 0.7-0.9) in those with larger brain volumes.

**Conclusions**: Larger brain volume quantified on clinical MRI of AIS patients at time of stroke purports a protective mechanism. The role of brain volume as a prognostic, protective biomarker has the potential to forge new areas of research and advance current knowledge of mechanisms of post-stroke recovery.

**Keywords**: 1. acute ischemic stroke, 2. Brain volume, 3. Outcome, 4. MRI

# Abbreviations

CCS: causative classification system

CE: major cardiac embolic

LAA: large artery atherosclerosis

mRS: modified Rankin Scale

NIHSS: National Institutes of Health Stroke Scale score

SAO: small artery occlusion

SD: standard deviation

IQR: interquartile range

$V_{brain}$: brain volume

WMHv: white matter hyperintensity volume

eR: effective reserve

AIS: acute ischemic stroke

MRI: magnetic resonance imaging

DWIv: acute stroke lesion volume segmented on diffusion-weighted imaging

AIC: Akaike Information Criterion

BIC: Bayes Information Criterion

# Introduction

Stroke is a leading cause of death [1] and disability [2] in adults. However, while early and accurate modeling of outcome holds great promise, the contributing factors to delineate and predict stroke outcome are poorly understood [3]. Brain volume ($V_{brain}$) has been studied in healthy and diseased populations [4–7], while brain reserve, the capacity of the brain to compensate for injury or disease, is often related to the maximum brain size in life, using intracranial volume as a surrogate measure [8]. Effective reserve (eR) assesses the remaining reserve after negative effects of insults and disease have been taken into account [9]. As such, eR is likely to relate to $V_{brain}$ at the time of scan. While $V_{brain}$ has been studied in relation to reserve in Alzheimer's disease [7], studies relating it to outcome after acute ischemic stroke (AIS) are missing.

Automated algorithms assess $V_{brain}$ using brain extraction and, if possible, tissue classification. However, algorithms are commonly designed for T1-weighted high-resolution images [10–12]. Recently, a translational methodology for clinical Fluid Attenuated Inversion Recovery (FLAIR) images, a common sequence in clinical assessment of AIS patients, has been developed [13].

Here, we posit $V_{brain}$ as a marker of brain health, which is influenced by the lifetime exposures to vascular risk factors. We hypothesized that larger $V_{brain}$ acts as a protective mechanism in AIS patients, leading to better post-stroke functional outcomes. To investigate this hypothesis, we assessed automatically determined $V_{brain}$ in 912 AIS patients and its association with outcome measured by the modified Rankin Scale (mRS) score, 60 to 190 days after stroke.

# Materials and Methods

## Standard protocol approvals, registrations, and patient consent

At time of enrollment, informed written consent was obtained from all participating patients or their surrogates. The use of human patients in this study was approved by each site's Institutional Review Board, in accordance with the 1964 Helsinki declaration and its later amendments or comparable ethical standards.

## Study design, setting, and patient population

This study was conducted between July 1, 2014 and March 16, 2019. The MRI-GENetic Interface Exploration (MRI-GENIE) study is a large-scale, international, hospital-based collaborative study of AIS patients [14]. Patients >18 years of age presenting to the Emergency Department of participating centers with signs and symptoms of AIS were approached for enrollment in the institutional stroke registries. Those with AIS confirmed on the brain MRI obtained within 48 hours of admission and blood samples available for genetic studies were eligible for inclusion in the MRI-GENIE study. Approval of the protocol was obtained from each site's institutional review board or ethics committee. FLAIR images, a standard sequence of each hospital's clinical AIS MRI protocol, were used for automated white matter hyperintensity volume (WMHv) extraction [13,15] in 2,528 patients from 12 sites (7 European, 5 US based). Basic demographics and clinical phenotypes, such as age, sex, prior stroke, hypertensive and smoking (ever/never) status, as well as stroke subtypes derived using the Causative Classification of Stroke system [16] (undetermined, large artery atherosclerosis (LAA), major cardiac embolic (CE-major), small artery occlusion (SAO) and 'other') were recorded on the hospital admission. Each patient was evaluated by a neurologist on admission for acute stroke severity measured using the National Institutes of Health Stroke Scale (NIHSS) score (0-42; 0 asymptomatic, 42: severe stroke). Functional outcome assessed as mRS scores (0-6; 0: asymptomatic; 6: death) was recorded between 60-190 days after stroke [17]. These data were available for a subset of 912 patients from six of the sites in the study. **Figure 1** describes the cohort selection and **Table 1** summarizes the data used in this work.

Additionally, we developed a validation set of 142 AIS patients from the larger cohort by deriving manual brain segmentations (**Table 1**). We used the $V_{brain}$ calculated from the manual brain extraction for quantitative evaluation of the efficacy of the brain extraction algorithm. Furthermore, acute stroke lesion volumes were manually segmented on diffusion-weighted imaging (DWIv) in 37.3% (340/912) of patients (median 2.83cc; interquartile range: 0.81-17.4cc), and utilized them to demonstrate that $V_{brain}$ estimates are independent of the acute stroke lesion size.

## Automated brain volume determination from FLAIR images

We used a brain extraction method for clinical FLAIR scans based on a U-Net convolutional neural network architecture (Neuron-BE), implemented using the open-source library neuron (http://github.com/adalca/neuron [18]). Neuron-BE contains five downsampling and five upsampling levels. Initially, the image intensities are roughly normalized by scaling the 97$^{th}$ percentile of the image intensities to 1. Each up-/downsampling is achieved using 2x2 upsample/maxpool operations and each level contains two convolution layers with 128 features per layer. Network parameters are optimized using the Adadelta stochastic optimizer [19] with mini-batches of size 16. We augmented the data to mimic the observed conditions in clinical data for each batch through random intensity scaling (contrast factor between 0.7-1.3), random ghosting effects (at most 3 "copies" of the brain), as well as additive Gaussian and Perlin noise (standard deviations of 0.4 and 0.5, respectively). The network parameters were learned on a set of 69 patient scans for which we manually outlined the brain. After applying our convolutional neural network to a patient's FLAIR image, we closed holes in the resulting segmentation mask, and identify the largest connected component as the brain mask [20]. These brain masks can be used to estimate total brain volume, i.e. ventricles, gray and white matter. Here, we applied the masks to the original scan and intensity normalized the resulting image using a mean-shift algorithm, setting the intensity of the normal appearing white matter to 0.75. From the resulting image voxel intensities below 0.375 were then excluded and $V_{brain}$ was calculated. A schematic of this process is shown in **Figure 2**.

## Statistical analysis and model description

To assess the efficacy of Neuron-BE, we calculated the correlation between total brain volume estimates in the validation set of 142 patients with the manually extracted volumes (Pearson's correlation coefficient). In addition, we calculated the overlap between manual and automated outlines (Dice coefficient) and confirmed that the estimated brain volume, after intensity normalization and thresholding, is independent of the DWIv in the set of 340 patients with manually extracted DWIv using Pearson's correlation coefficient.

**Brain volume model.** We then determined the collinearity between the phenotypes used in our study and assessed associations with $V_{brain}$ using a generalized linear model with backward elimination, iteratively removing non-significant variables. The *initial brain volume model* is given by

$$V_{brain} \sim age + sex + \log(WMHv) + prior\ stroke + hypertension + smoking + NIHSS + diabetes + CE +$$
$$SAO + LAA + Other, \textbf{(Eq.1)}$$

including age, sex, WMHv, history of prior stroke, CCS stroke subtypes CE, SAO, LAA, and Other (yes/no), NIHSS, diabetes, hypertensive and smoking status. The CCS category 'Undetermined' was not included. Prior to model fit, WMHv were log transformed and continuous variables (age, WMHv, and brain volume) were standardized by subtracting the mean and dividing by their standard deviations.

**Outcome model.** We used the determined relationships between observed phenotypes and the extracted brain volume determined in our brain volume model to guide the selection of interaction terms to be included in the *outcome model*, given by

$$mRS \sim age + sex + \log(WMHv) + prior\ stroke + hypertension + smoking + NIHSS + diabetes + CE +$$
$$SAO + LAA + Other + V_{brain} + \sum interaction\ terms. \textbf{(Eq. 2)}$$

Parameters of the outcome model were estimated using ordinal regression and backward elimination was performed, iteratively removing non-significant parameters. We investigated the efficacy of this model by comparing it to a *base model* using a $\chi^2$ test, the Akaike Information Criterion (AIC) and the Bayes Information Criterion (BIC). The base model is defined as the outcome model after backward elimination without $V_{brain}$. We furthermore included site, i.e. institution of acquisition, as a nuisance parameter in all models.

To validate our findings, we utilized 100 5-fold cross-validation, where the data were divided 100 times into five disjoint sets/folds of approximately equal size. The model fit was then repeated five times by leaving one of the folds out for each set of folds, which allowed us to assess the stability of our parameter estimates by calculating mean and standard deviation. All statistical analyses were conducted using the computing environment R [21]. Significance was set at $P<.05$.

## Data Availability Statement

The authors agree to make the data, methods used in the analysis, and materials used to conduct the research available to any researcher for the express purpose of reproducing the results and with the explicit permission for data sharing by the local institutional review board.

# Results

An overview of the cohort characteristics is found in Table 1. Of the 912 stroke patients with an average age (standard deviation, SD) of 65.3 (14.5) years, 58.3% were male, 65.2% were hypertensive, and 56.9% smoked. On average, patients had a median mRS score of 2 (inter-quartile range (IQR): 2; mRS of 0: 17%; mRS of 1: 37%; mRS of 2: 19%; mRS of 3: 13%; mRS of 4: 9%; mRS of 5: 1%), and an NIHSS score (mean (SD)) of 5.3 (5.5), while 7.6% have had a prior stroke. Significant difference in patients with manual stroke lesion volume estimates were found for smoking (higher prevalence), mRS (better outcome), and SAO (lower prevalence) and Undetermined (higher prevalence) stroke subtypes, compared to those where manual stroke lesion estimates were not available.

The Pearson's correlation coefficient $r$ and mean Dice coefficient $d$ between manual and automated total brain volumes in the validation set were determined as r=0.94 and d=0.94±0.02. Additionally, the correlation between DWIv and $V_{brain}$ was not significant, whereas $V_{brain}$ was associated with age, sex, hypertension, WMHv, and smoking. Univariable analysis results are summarized in **Table 2**.

**Brain volume model.** We reduced the full *brain volume model* (**Eq. 1**) through backward elimination (see Table 3). Finally, we obtained associations of $V_{brain}$ with age, sex, and prior stroke. The model estimation is summarized in

**Table 2**. Our results showed that male patients have larger $V_{brain}$ compared to female patients. Prior stroke, as well as an increase in age (per 14.4 years in the model) were associated with a decreased brain volume (per 155.1cc).

**Outcome model.** The *full outcome model* (**Eq. 2**), with interaction terms between $V_{brain}$ and age, sex, and prior stroke, was then reduced using backward elimination (see Table 3). In the final step, we obtained associations of mRS with age, prior stroke, hypertension, NIHSS, LAA and Other stroke subtypes, as well as $V_{brain}$.

The parameters of the *outcome* and *base models*, i.e. the outcome model without $V_{brain}$ as a factor, are summarized in **Table 2**. Older individuals, those with history of prior stroke, HTN, those with LAA and Other stroke subtype, and with higher NIHSS Score on admission had worse outcome (higher mRS). The inverse was observed for $V_{brain}$. Comparison of both models suggests that including $V_{brain}$ in the model explains the observed data better than the corresponding base model ($AIC_{base}$=2714.9; $AIC_{outcome}$=2703.8; $BIC_{base}$=2796.8; $BIC_{outcome}$=2790.4).

Odds ratios (OR) and confidence intervals (CI) of both models (**Figure 3**) showed an increase in odds of worse outcome (higher mRS) for age (base OR: 1.41; outcome OR: 1.32), NIHSS (base OR: 1.19; outcome OR: 1.19), hypertension (base OR: 1.42; outcome OR: 1.38), prior stroke (base OR: 2.13; outcome OR: 2.00), and in case of an LAA (base OR: 1.37; outcome OR:1.37), and 'other' stroke subtype (base OR: 1.93; outcome OR: 1.84). In the outcome model, an increase in $V_{brain}$ showed a reduction in odds for worse outcome (OR: 0.79). Validating our results using cross-validation showed consistent trends with ORs of 1.24±0.04 for age, 2.27±0.29 for prior stroke, 1.44±0.09 for hypertension, 1.35±0.10 for CCS-LAA, 1.66±0.17 for CCS-Other, 1.19±0.01 for NIHSS and 0.79±0.02 for $V_{brain}$. Odds-ratios for age and $V_{brain}$ reflect an increase by 14.4 years and 155.1cc, respectively, due to the standardization used in the model fit.

# Discussion

In a large cohort of AIS patients, we demonstrated that $V_{brain}$ is an important factor of post-stroke outcome. We derived the $V_{brain}$ estimates automatically on clinical FLAIR MRI sequences using a novel, deep-learning algorithm, and confirmed good agreement of the automated total brain volumes with manually extracted "gold standard" estimates. Furthermore, we verified independence of $V_{brain}$ estimates from measurement of acute infarct volume (further strengthened by lack of association between $V_{brain}$ and NIHSS) and demonstrated that a larger $V_{brain}$ acts as a protective factor in functional outcome after stroke.

Utilizing the extracted $V_{brain}$ as an independent variable in the *outcome model* of acute stroke, we demonstrated that age, history of prior stroke and $V_{brain}$ are important markers of functional post-stroke outcome. Both age and history of prior stroke have been reported previously; however, the link between $V_{brain}$ and mRS is novel. Moreover, these findings are significant, as they bridge a long-standing conceptual divide between the burden of pre-existing brain pathology and its impact on susceptibility of the brain to acute insult such as stroke. Larger brain volumes have been associated with the notion of higher brain reserve, in which intracranial volume is used as a measure of maximum brain volume over the life span [9,22]. However, the brain volume calculated here does not reflect each patient's maximum brain volume, but the current volume which may have been altered due to disease or normal brain aging. Brain volume at the time of MRI scan directly relates to the recently proposed concept of effective reserve (eR), which quantifies how the remaining brain reserve compensates for new diseases or insults [9]. This is further reflected in the protective nature of larger brain volumes, which were associated with better outcome in our study, and may represent a sign of eR in AIS populations.

Brain volume, however, is modified by common stroke risk factors, as shown in the *brain volume model*. Using backward elimination, our analysis of brain volume determinants reproduced known associations in healthy populations, where increasing age leads to a reduced brain volume due to general brain atrophy [8] and males having larger brain volumes compared to women, likely due to anthropometric differences between sexes [23]. Additionally, we show a reduction in $V_{brain}$ depending on history of prior stroke, which may indicate increased atrophy after incidence of stroke, independent of aging. Furthermore, we observed a negative trend of $V_{brain}$ with hypertension. Although this analysis did not reach statistical significance ($p<0.1$), associations of hypertension with brain volume have been previously reported [24]. We were not able to identify an effect of WMHv on $V_{brain}$ in our study. Many of the presented associations are thought to cause increased WMHv; however, the increase of WMHv may be independent from alterations in $V_{brain}$. Future studies will aim to investigate whether WMHv may be associated with local alterations to the specific brain compartments, such as periventricular areas.

This study has several limitations. It should be emphasized that this is not a direct study of brain atrophy, as intracranial volume or serial imaging was not available for this cohort. Instead, this study focuses on the brain volume, which may act as a more direct measure of eR, or a cross-sectional metric of "lifetime brain health," assessed at the time of stroke. Other factors can influence the estimation of brain volume in this study, such as brain swelling, utilization of FLAIR imaging instead of commonly used T1 sequence, as well as the clinical resolution of the investigated data. However, the comparison of the $V_{brain}$ estimates of the presented method to manually determined brain volumes showed good agreement, while including site as a nuisance factor in all models, and no associations were found between acute lesion size and total brain volume. In addition, the time of day at acquisition has been shown to influence brain volume with larger brain volumes earlier in the day. However, Nakamura et al. [25] estimated the error on brain volume estimation to be below 0.2% at a resolution of 1x1x3mm$^3$, in our cohort corresponding to 2-4cc of brain volume, and subsequently much smaller than the increment of 155.1cc described in our analysis. Additionally, our outcome model did not take acute stroke lesion volume, its location, or other chronic discrete infarction beyond WMHv into account, because only a small subset (n=340; 37%) of our patients had DWIv available for analysis. However, we utilized the NIHSS score as a measure of stroke severity, and given its collinearity with DWIv, we do not expect any significant changes in the model with respect to its effect on $V_{brain}$. Similarly, there was no interaction between CCS stroke subtype and $V_{brain}$. Furthermore, with regard to stroke location, it is unlikely that the location, independent of lesion size, will affect the estimates of $V_{brain}$. To identify chronic lesions, we utilized a fully automated segmentation pipeline specifically developed for clinical FLAIR sequences, while excluding hyperintensities not being classified as leukariosis. While smaller, predominantly subcortical strokes could have been missed or perhaps erroneously included as part of the WMHv in case of small chronic lacunar infarctions, to date not dedicated pipeline for identifying chronic, discrete infarctions in mono-modal, clinical data, other than leukariosis, exist. Future developments of such pipelines may further help to identify additional associations between $V_{brain}$ and chronic lesion burden.

The goal of our study was to assess whether antecedent brain volume is an independent factor that can be utilized in stroke outcome models and we have demonstrated its independence with respect to the acute lesion volume and etiologic stroke subtype. Additionally, other factors, such as stroke treatment can influence the overall outcome of a patient. While we had no information on acute interventions in this cohort, the treatment itself is not likely to systematically influence brain volume or to have major effects in the acute stages of the stroke (<48 hours) in which $V_{brain}$ was assessed.

In this work, we present a simple, but important, additional biomarker for assessment of functional post-stroke outcome. The proposed brain extraction methodology can provide consistent, precise and reliable estimates of brain volume on clinical brain MRI scans without time intensive manual labor. Using cross-validation and large-scale, international, multi-site data, we demonstrate that our results are generalizable and are directly applicable to clinical grade data, making brain volume a complementary marker for state-of-the-art stroke outcome assessment in the clinic.

# Conclusion

We established brain volume as an important, but often overlooked, biomarker associated with AIS outcome. We showed that brain volume is associated with age, sex and history of prior stroke and demonstrated that it has an independent, protective effect, where a larger brain volume relates to better outcome. Importantly, the assessment of brain volume on clinical MR images offers an opportunity for immediate translation and opens up new areas of research to help advance current knowledge of risks and outcomes in AIS populations.

**Acknowledgements:**
This project has received funding from the European Union's Horizon 2020 research and innovation programme under the Marie Sklodowska-Curie grant agreement No 753896. Major support for this study was provided by the MRI-GENIE study (NIH-NINDS R01NS086905, Rost PI), the Swedish Research Council (C.J.), the Swedish Heart and Lung Foundation (C.J.), and the Swedish State under the ALF agreement (C.J.).

**Disclosures:**
Dr. Schirmer is in part supported by Horizon 2020 grant agreement No 753896.
Ms. Donahue reports no disclosures.
Mr. Nardin reports no disclosures.
Dr. Dalca is supported by NIH 1R21AG050122.
Dr. Giese reports no disclosures.
Dr. Etherton is supported in part by the AHA Clinical Scientist Training Program (17CPOST33680102).
Mr. Mocking reports no disclosures.
Ms. McIntosh reports no disclosures.
Dr. Cole is partially supported by National Institutes of Health (NIH) grants R01-NS100178 and R01-NS105150, the US Department of Veterans Affairs, the American Heart Association (AHA) Cardiovascular Genome-Phenome Study (Grant-15GPSPG23770000), and the AHA-Bayer Discovery Grant (Grant-17IBDG33700328).
Dr. Holmegaard reports no disclosures.
Dr. Jood reports no disclosures.

Dr. Jimenez-Conde in part supported by Fondos de Investigación Sanitaria ISC III (PI10/02064), (PI15/00451); and Fondos FEDER/EDRF Spanish stroke research network INVICTUS+ (RD16/0019/0021).

Dr Kittner is partially supported by NIH grant, R01 NS086905-01.

Dr. Lemmens is a senior clinical investigator of FWO Flanders.

Dr. Meschia is co-Principal Investigator for the CREST-2 trial (U01 NS080168) and receives some support for the CREST-H study (R01 NS097876). He also serves as chair for the NeuroNEXT DSMB.

Dr. Rosand is supported by grants from the US National Institutes of Health and the OneMind Foundation and the Henry and Allison McCance Center for Brain Health.

Dr. Roquer is in part supported by Spain's Ministry of Health (Ministerio de Sanidad y Consumo, Instituto de Salud Carlos III FEDER, RD12/0042/0020).

Dr. Rundek is supported by the grants from the NINDS R37 NS 29993; R01NS040807; and NIH/NCATS Miami Clinical and Translational Science Institute U54TR002736/ KL2TR002737.

Dr. Sacco is supported by the grants from the NINDS R37 NS 29993; R01NS040807; and NIH/NCATS Miami Clinical and Translational Science Institute U54TR002736/ KL2TR002737.

Dr. Schmidt reports no disclosures.

Dr. Sharma reports no disclosures.

Dr. Slowik reports no disclosures.

Dr. Stanne reports no disclosures.

Dr. Vagal is supported by R01 NINDS NS100417, R01 NINDS NS30678, NIH/NINDS 1U01NS100699, research grant Cerovenus, GE Healthcare research grant.

Dr. Wasselius reports no disclosures.

Dr. Woo is in part supported by NIH funds.

Dr. Bevan reports no disclosures.

Dr. Heitsch reports no disclosures.

Dr. Phuah reports no disclosures.

Dr. Strbian received funding from the Helsinki University Central Hospital governmental subsidiary funds for clinical research.

Dr. Tatlisumak reports grants from Helsinki University Central Hospital, grants from University of Gothenburg, grants from Sahlgrenska University Hospital, grants from Sigrid Juselius Foundation, during the conduct of the study; grants and personal fees from Boehringer Ingelheim, personal fees from Bayer, grants from BrainsGate, grants from Bayer, grants from Pfizer, personal fees from Lumosa Pharm, grants from Portola Pharm, outside the submitted work; In addition, Dr. Tatlisumak has a patent use of a mast cell activation or degranulation blocking agent in the manufacture of a medicament for the treatment of a patient subjected to thrombolyses. Patent no: US8163734. Filed: February 13, 2004. Issued: April 24, 2012.

Mr. Levi reports no disclosures.

Dr. Attia reports no disclosures.

Dr. McArdle is in part supported by NIH NINDS (R01 NS100178, R01 NS105150).

Dr. Worrall is in part supported by U10 NS086513/U24 NS107222 and U01 NS069208. Dr. Worrall is the Deputy Editor of the journal Neurology.

Dr. Wu is in part supported by NIH (P50NS051343, R01NS082285, R01NS086905) and National Institute of Biomedical Imaging and Bioengineering (P41EB015896).

Dr. Jern reports no disclosures.

Dr. Lindgren is supported by Region Skåne, Lund University, the Swedish Heart and Lung Foundation, the Freemasons Lodge of Instruction Eos Lund, Skåne University Hospital, the Foundation of Färs&Frosta—one of Sparbanken Skåne's ownership Foundations, and the Swedish Stroke Association.

Dr. Maguire reports no disclosures.

Dr. Thijs has been a consultant for Bayer, Boehringer Ingelheim, BMS/Pfizer and Amgen; is in part supported by the Victorian Government and in particular the funding from the Operational Infrastructure Support Grant.

Dr. Rost is in part supported by NIH-NINDS (R01NS086905 and R01NS082285).

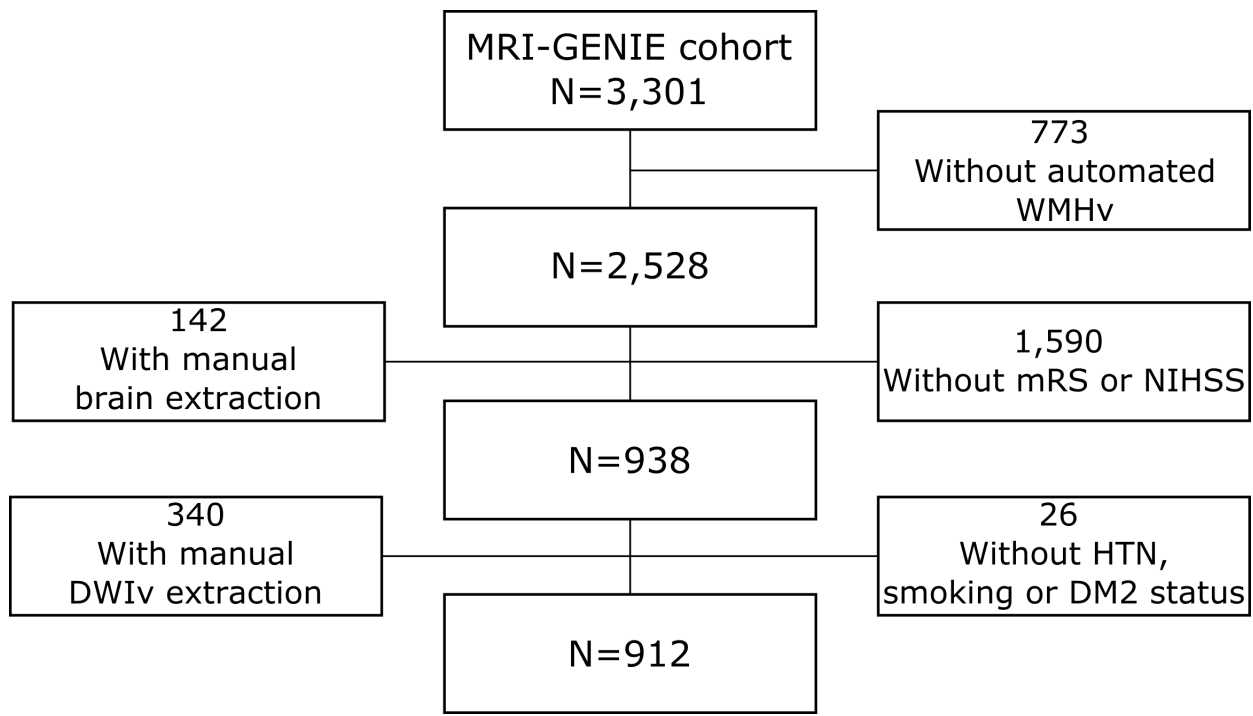

**Figure 1: Cohort selection**. Of the patients with automated WMHv estimates, 142 underwent manual brain extraction (validation set). A subset of 912 patients was used for analysis in this study, of which 340 had manually determined acute lesion volume (DWIv), which form the DWIv set. *Abbreviations:* DWIv - diffusion weighted imaging acute lesion volume, mRS - modified Rankin Scale score, WMHv - white matter hyperintensity volume, HTN - hypertension, DM2 - type 2 diabetes.

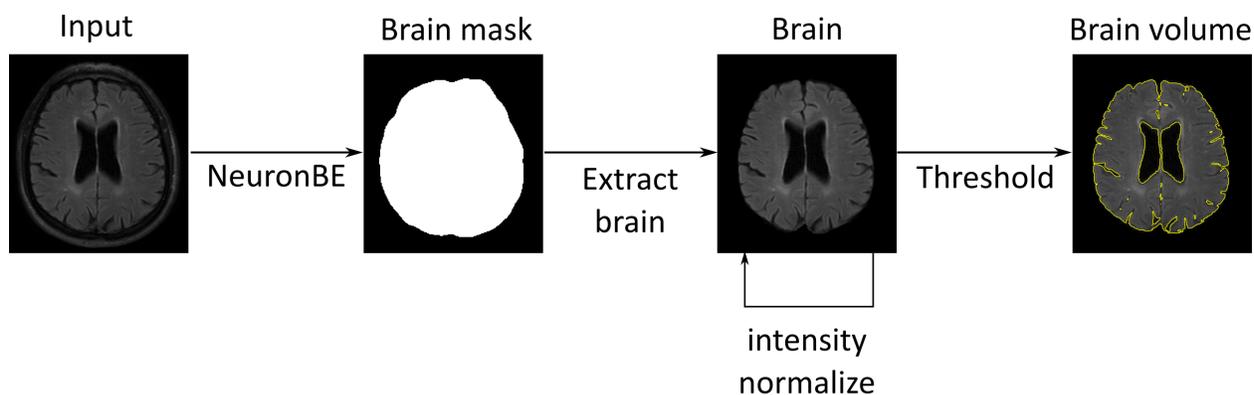

**Figure 2: Pipeline schematic to extract brain volume.** After brain extraction utilizing Neuron-BE, image intensities were normalized and then thresholded to determine a final brain mask (yellow outline) from which brain volume was calculated.

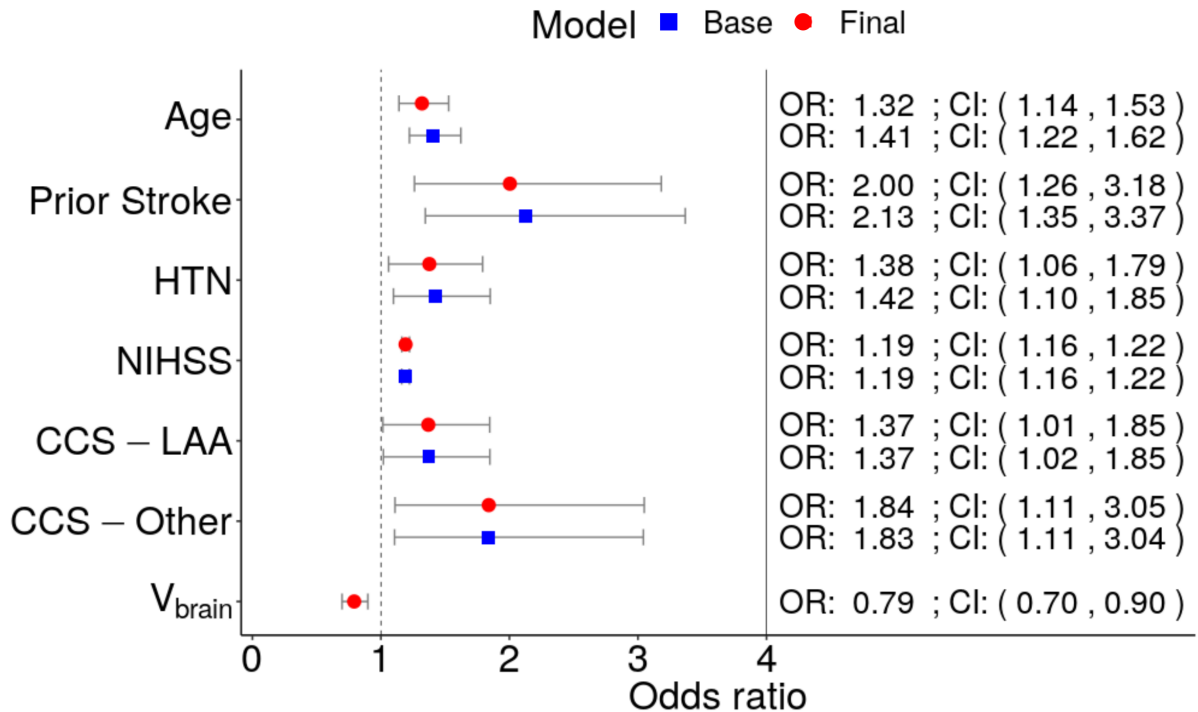

**Figure 3: Odds ratios estimated for base (blue) and final (red) model of outcome (mRS score 0-6).** In both models, age and history of stroke increase the odds of a higher mRS score, reflecting worse outcome. Larger brain volumes reduce the odds of a higher mRS score, reflecting better outcome. The odds ratios for age and $V_{brain}$ reflect an increase by 14.4 years and 155.1cm$^3$, respectively, due to the standardization used in the model fit.

**Table 1: Study cohort characterization**. Smoking was assessed as ever/never, based on past or current history of smoking. *Abbreviations*: CCS: causative classification system; CE major: major cardiac embolic; LAA: large artery atherosclerosis; mRS: modified Rankin Scale; NIHSS: National Institutes of Health Stroke Scale score; SAO: small artery occlusion; sd: standard deviation; IQR: interquartile range; $V_{brain}$: brain volume; WMHv: white matter hyperintensity volume.

| | Overall | Validation set | DWIv set |
|---|---|---|---|
| **n** | 912 | 142 | 340 |
| **Age, years (mean (sd))** | 65.3 (14.5) | 62.1 (16.9) | 66.52 (14.1) |
| **Male sex, n (%)** | 532 (58.3) | 86 (59.7) | 202 (59.4) |
| **Smoking, n (%; n missing)** | 519 (56.9) | 75 (53.6; 4) | 212 (62.4) |
| **Hypertension, n (%; n missing)** | 595 (65.2) | 91 (64.5; 3) | 207 (60.9) |
| **Prior stroke, n (%; n missing)** | 70 (7.6) | 25 (18.2; 1) | 20 (5.9) |
| **mRS score (median (IQR); n missing)** | 2 (2) | 1 (2; 99) | 1 (1) |
| **NIHSS score (mean (sd); n missing)** | 5.3 (5.5) | 4.8 (4.9; 84) | 5.3 (5.5) |
| **WMHv, cm³; mean (sd)** | 11.0 (12.9) | 14.2 (16.7) | 10.7 (13.0) |
| **$V_{brain}$, cm³; mean (sd)** | 1403.1 (155.1) | 1273.7 (164.5) | 1392.9 (154.9) |
| **CCS subtype** | | | |
| **Undetermined, n (%)** | 349 (38.3) | 61 (42.4) | 149 (43.8) |
| **CE Major, n (%)** | 162 (17.8) | 23 (16.0) | 58 (17.1) |
| **LAA, n (%)** | 194 (21.3) | 30 (20.8) | 71 (20.9) |
| **SAO, n (%)** | 149 (16.3) | 20 (13.9) | 39 (11.5) |
| **Other, n (%)** | 58 (6.4) | 10 (6.9) | 23 (6.9) |

**Table 2: Parameter estimates for all models**. 95% confidence intervals are reported in parentheses. *Abbreviations*: HTN: hypertension; CE major: major cardioembolic; Eq: equation; LAA: large artery atherosclerosis; mRS: modified Rankin Scale; NIHSS: National Institutes of Health Stroke Scale score; SAO: small artery occlusion; $V_{brain}$: brain volume; WMHv: white matter hyperintensity volume

| Model | Brain volume (Linear regression) | | | | Outcome (mRS; Ordinal regression) | | | | | |
| | Univariable | | Multivariable (Eq. 3) | | Univariable | | Multivariable (Eq. 5) | | Base (Eq. 6) | |
| | Estimate | *P* | Estimate | *P* | Estimate | *P* | Estimate | *P* | Estimate | *P* |
|---|---|---|---|---|---|---|---|---|---|---|
| **Age** | -0.33 (-0.39--0.26) | <.001 | -0.30 (-0.36--0.25) | <.001 | 0.33 (0.21-0.45) | <.001 | 0.28 (0.13-0.42) | <.001 | 0.34 (0.20-0.48) | <.001 |
| **Sex (Male)** | -0.60 (-0.69--0.52) | .03 | 1.00 (0.89-1.11) | <.001 | -0.51 (-0.75--0.27) | <.001 | - | - | - | - |
| **Prior stroke** | 0.03 (-0.04-0.10) | .11 | -0.21 (-0.41-0.00) | .04 | 1.12 (0.69-1.54) | <.001 | 0.70 (0.23-1.16) | .001 | 0.76 (0.30-1.21) | .003 |
| **HTN** | 0.23 (0.12-0.33) | .04 | - | - | 0.51 (0.26-0.76) | <.001 | 0.32 (0.06-0.58) | .008 | 0.35 (0.09-0.62) | .01 |
| **NIHSS** | -0.00 (-0.01-0.00) | .31 | - | - | 0.17 (0.15-0.19) | <.001 | 0.18 (0.15-0.20) | <.001 | 0.18 (0.15-0.20) | <.001 |
| **LAA** | 0.01 (-0.06-0.09) | .07 | - | - | 0.59 (0.30-0.88) | <.001 | 0.31 (0.01-0.61) | .03 | 0.32 (0.02-0.61) | .04 |
| **Other** | -0.02 (-0.09-0.04) | .13 | - | - | -0.19 (-0.65-0.27) | .41 | 0.61 (0.10-1.11) | .01 | 0.61 (0.10-1.11) | .01 |
| **SAO** | 0.01 (0-0.06-0.08) | .08 | - | - | -0.05 (-0.35-0.25) | .75 | - | - | - | - |
| **CE** | 0.04 (-0.03-0.11) | .07 | - | - | 0.37 (0.05-0.69) | .02 | - | - | - | - |
| **WMHv** | -0.09 (-0.15--0.02) | .009 | - | - | 0.04 (-0.08-0.16) | .49 | - | - | - | - |
| **Smoking** | -0.08 (-0.18-0.02) | .04 | - | - | -0.07 (-0.31-0.16) | .55 | - | - | - | - |
| **Diabetes** | 0.03 (-0.04-0.10) | .07 | - | - | 0.63 (0.35-0.91) | <.001 | - | - | - | - |
| **$V_{brain}$** | - | - | - | - | -0.34 (-0.46--0.22) | <.001 | -0.23 (-0.36--0.11) | <.001 | - | - |

**Table 3: Summary of backward elimination steps.** Phenotypic variables removed during each iterative step, including the corresponding p-values in parentheses, for both base and outcome models. Interaction terms are represented by ':'.

| Model | Iteration | | | | | | | | |
|---|---|---|---|---|---|---|---|---|---|
| | 1 | 2 | 3 | 4 | 5 | 6 | 7 | 8 | 9 |
| **Base** | SAO (.80) | Other (.63) | CE (.63) | NIHSS (.50) | WMHv (.40) | LAA (.41) | Diabetes (.34) | Smoking (.24) | Hypertension (.08) |
| **Outcome** | CE (.96) | WMHv (.92) | Sex:$V_{brain}$ (.78) | Prior stroke:$V_{brain}$ (.71) | Smoking (.44) | SAO (.30) | Age:$V_{brain}$ (.08) | Sex (.07) | - |